\documentclass[%
 aip,
 sd,%
 amsmath,amssymb,
 reprint,%
]{revtex4-1}

\usepackage{graphicx}
\usepackage{dcolumn}
\usepackage{bm}

\begin{document}

\preprint{AIP/123-QED}

\title{The molecular structure of the interface between water and a hydrophobic substrate is liquid-vapor like}

\author{Adam P. Willard}
\affiliation{Department of Chemistry, Massachusetts Institute of Technology.}
\email{awillard@mit.edu}

\author{David Chandler}%
\affiliation{ Department of Chemistry, University of California, Berkeley.}

\date{\today}

\begin{abstract}
With molecular simulation for water and a tunable hydrophobic substrate, we apply the instantaneous interface construction [A. P. Willard and D. Chandler, J. Phys. Chem. B \textbf{114} 1954 (2010)] to examine the similarity between a water-vapor interface and a water-hydrophobic surface interface.  The intrinsic interface refers to molecular structure in terms of distances from the instantaneous interface. We show that attractive interactions between a hydrophobic surface and water affect capillary wave fluctuations of the instantaneous liquid interface, but these attractive interactions have essentially no effect on the intrinsic interface. Further, the intrinsic interface of liquid water and a hydrophobic substrate differs little from that of water and its vapor.The same is not true, we show, for an interface between water and a hydrophilic substrate.  In that case, strong directional substrate-water interactions disrupt the liquid-vapor-like interfacial hydrogen bonding network. 
\end{abstract}

\pacs{Valid PACS appear here}
\maketitle

\section{Introduction}

More than forty years ago, Frank Stillinger~\cite{FS73} argued that the liquid water interface adjacent to extended non polar hydrophobic substrate is universal and shares the same microscopic features as a liquid-vapor interface. Subsequent theory and molecular simulations support this idea~\cite{LCW,DC05,PT08} .  On the other hand, some simulation data shows that interfaces between liquid water and physically realistic hydrophobic substrates lack discernible vapor (or dewetting) regions. Instead, water density profiles extending from realistic models of typical hydrophobic substrates can exhibit molecular layering, the details of which depend on the chemical identity of the substrate~ \cite{MEP01,DC02,BJB04,MP05,LM07,SG09,SG11,NC12}.  Here, we demonstrate that this substrate dependence reflects significant sensitivity of the position of the average interface to weak adhesive forces.  At the same time, the fluctuations of the interface, and the mean molecular structure relative to the instantaneous interface~\cite{APW10} are both insensitive to weak adhesive forces. We refer to this dynamic frame of reference in which molecular structure is resolved in terms of distances from the time-varying position of the instantaneous interface as the \textit{intrinsic} interface.  For water adjacent to an extended hydrophobic surface, the structure of an intrinsic interface exhibits almost no substrate dependence, and  this intrinsic interface is quantitatively similar to that of the water-vapor interface. This commonality is not present at the interfaces between water and hydrophilic substrates, which we also demonstrate in this paper. The results presented in this paper thus constitute explicit confirmation of Stillinger's 1973 hypothesis, and they reconcile the apparent inconstancy presented by some numerical studies.

In our classical molecular dynamics simulations, the general system consists of a slab of liquid water in contact with a model substrate.  It is simultaneously in equilibrium with its vapor phase and in contact with a substrate.  See Fig.~\ref{fig:system}a.  An adjustable water-substrate attraction controls the hydrophobicity of the substrate. The water-substrate attractive interactions are isotropic with respect to molecular orientation and weak compared to hydrogen bonding.  They thus influence the position interfacial molecules without necessarily disrupting the hydrogen bonding structure of the liquid interface. The methodology we employ involves separating the collective fluctuations of soft liquid interfaces from the intrinsic molecular structure. This separation requires identifying the time-varying position of the liquid-water instantaneous interface, which we accomplish following the algorithm described in Ref.~\cite{APW10}. We use the instantaneous interface as a dynamic frame of reference, performing a spatial transformation that defines the vertical position of each water molecule relative to the local instantaneous interface rather than a fixed Cartesian plane. By doing so the spatial deformations of the liquid phase boundary are projected out. The degrees of freedom that remain after this transformation constitute our definition of the intrinsic interface.

Our methods are described in the next section.  After that, we present and discuss our results. 
 
\section{Methods}
\subsection{Simulation Details}
\begin{figure*}
\centering
  \includegraphics[width=6.69in]{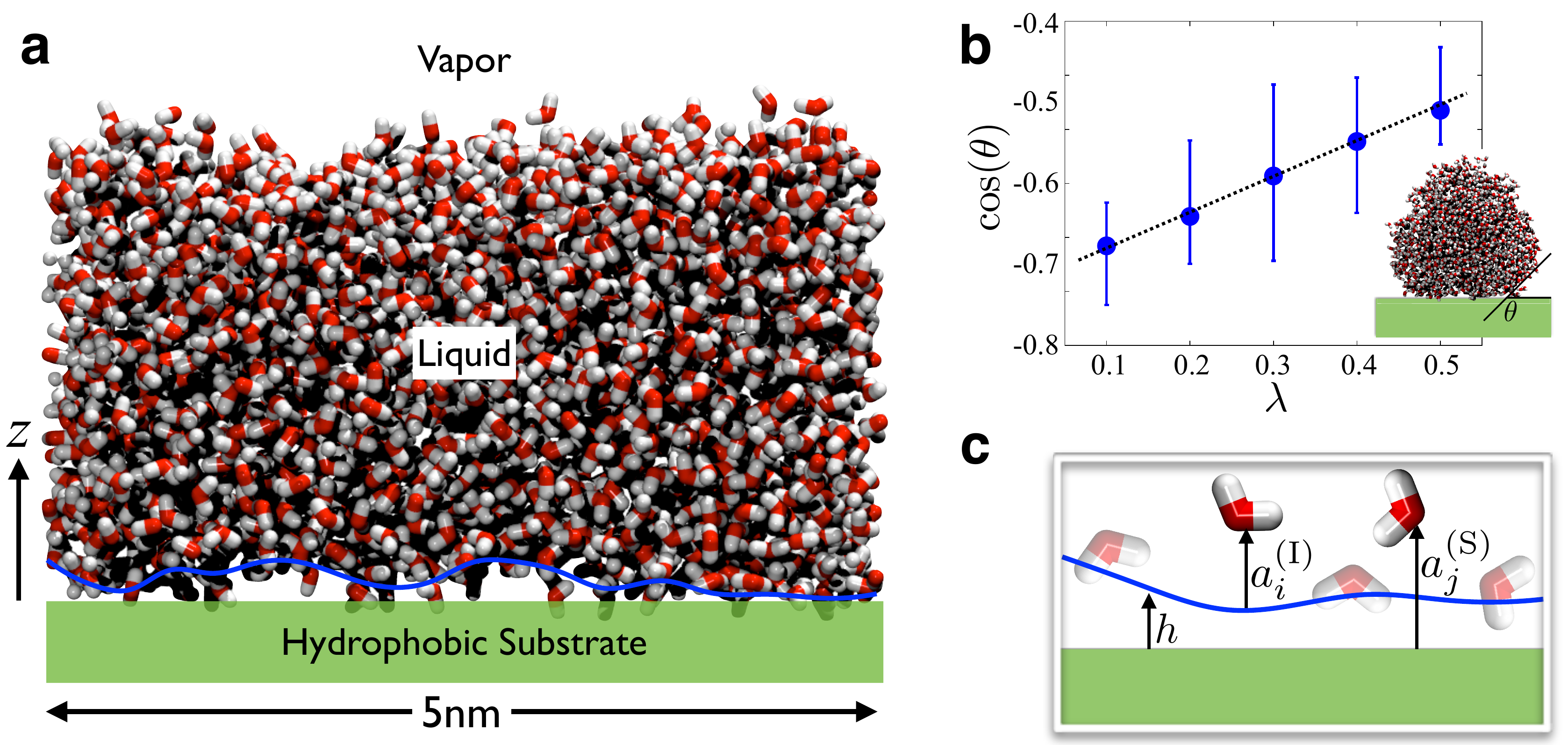}
  \caption{(a) A snapshot of the simulation system. Water molecules are rendered in red and white and the hydrophobic interface is rendered in green.  The position of the instantaneous interface is represented by a solid blue line. (b) The dependence of contact angle, computed using simulation data, on the substrate-water attractive parameter $\lambda$. (c) Schematic illustration of the coordinate system for the standard and intrinsic interface. The vertical position of a molecule $j$ relative to the standard interface is $a^{(\mathrm{S})}_j$ and the vertical position of a molecule $i$ relative to the intrinsic interface is $a^{(\mathrm{I})}_i$. }
  \label{fig:system}
\end{figure*}

The model system consists of a slab of 2261 SPC/E water molecules~\cite{SPCE} in periodically replicated simulation cell measuring $5 \times 5 \times 10~\mathrm{nm}^3$ in the $x$-,  $y$- and $z$-direction, respectively.  It is propagated in time using standard molecular dynamics at a temperature of 298K. A rendering of the simulation cell is shown in Fig.~\ref{fig:system}. The simulation cell is long enough in the $z-$dimension so that the water condenses against the substrate and forms both a water-substrate and a water-vapor interface. Although the overall simulation cell is held at constant volume, the presence of the free water-vapor interface acts as a natural barostat to the liquid. At the bottom of the simulation cell extending across the $x$-$y$ plane is a planar hydrophobic substrate whose interactions with individual water molecules is of the form,
\begin{equation}
w_\lambda(z_i) = w_0(z_i) + \lambda w_1(z_i).
\label{eq:wca_pot}
\end{equation}
where $z_i$ is the position of the center of the oxygen atom of the $i$th water molecule, and the functions $w_0(z)$ and $w_1(z)$ are the repulsive WCA potential~\cite{WCA}, and the attractive branch of the Lennard-Jones potential respectively. The two parts of the water-substrate potential are given by,
$$
  	w_0(z) = \left\{ \begin{array}{ll}
		4 \epsilon_\mathrm{s} \left [ (\sigma_\mathrm{s}/z)^{12} - (\sigma_\mathrm{s}/z)^6 + 1/4 \right ] , & \quad   z \leq 2^{1/6} \sigma_\mathrm{s}, \\
          	0, & \quad z > 2^{1/6} \sigma_\mathrm{s}, \end{array} \right.
$$
and
$$
  	w_1(z) = \left\{ \begin{array}{ll}
		-\epsilon_\mathrm{s}, & \quad z \leq 2^{1/6} \sigma_\mathrm{s}, \\
          	4 \epsilon_\mathrm{s} \left [ (\sigma_\mathrm{s}/z)^{12} - (\sigma_\mathrm{s}/z)^6\right ], & \quad z > 2^{1/6} \sigma_\mathrm{s}, \end{array} \right.
$$
where $\sigma_\mathrm{s}=5\mathrm{\AA}$, $\epsilon = 1.825\,k_\mathrm{B}T$, and the quantity $\lambda$ tunes the strength of the water-substrate attraction. We consider a range of values for $\lambda$ between 0.1 and 0.5, with $\lambda=0.3$ being approximately equal to the effective potential between water and a surface composed of alkane chains. Averages were generated using 6000 snapshots equally spaced over a $750\,\mathrm{ps}$ simulation.

To measure contact angles, 1387 SPC/E water molecules were placed inside a much larger simulation cell (see Fig.~\ref{fig:system}b inset), one that measured $15 \times 15 \times 15~\mathrm{nm}^3$. In this larger cell the number of water molecules is insufficient to bridge the periodic boundaries and responds by forming a liquid droplet. Contact angles were estimated by extrapolation on the horizontal center of mass corrected mean solvent density profile.

To generate a model hydrophilic substrate a plane was drawn through a slab of liquid water that had been equilibrated at 298K.  The positions of all the molecules whose oxygen atoms reside on one side of the plane were frozen in space to produce the hydrophilic substrate~\cite{AJP10}. 

\subsection{Instantaneous interface and relative coordinates}
We refer to the ``standard'' interface to indicate a Cartesian frame of reference and the ``intrinsic'' interface to indicate an instantaneous interface frame of reference. To generate the latter we utilize the construction presented in Ref.~\cite{APW10} for identifying the time-varying position of the instantaneous water interface. The procedure associates a Gaussian density function with the discrete position of each water molecule in the system. The width of the Gaussian is then a coarse-graining length.  Here, we use 2.4 $\mathrm{\AA}$ as the width, which is approximately the molecular diameter. The coarse-grained density field is the sum over at the Gaussian density functions.  For an individual snapshot of the system the position of the instantaneous interface is the set of points on the coarse-grained density field whose value is equal to a density value intermediate between the average density of the bulk liquid and the bulk vapor.  Here, we take an intermediate value as one-half that of the bulk liquid, $\rho_\ell$.  Any choices of coarse-graining lengths between 2.2 $\mathrm{\AA}$ and 3.5 $\mathrm{\AA}$, and any choice of intermediate densities between 0.3 $\rho_\ell$ and 0.7 $\rho_\ell$ will yield results essentially identical to those presented below. 

We utilize two specific measures of molecular structure in order to characterize the interface between water and a variety of hydrophobic or hydrophilic substrates. One is the mean solvent density projected along an axis perpendicular to the substrate surface, given by
\begin{equation}
	\rho^{(\alpha)}(z) = \frac{1}{A}\left \langle \sum_{i=1}^{N_\mathrm{w}} \delta(a_i^{(\alpha)}-z) \right \rangle,
	\label{eq:1}
\end{equation}		
where the superscript $\alpha$ indicates the relative coordinate system ($\alpha=\mathrm{S}$ for the standard interface and $\alpha=\mathrm{I}$ for the intrinsic interface), $A$ is the substrate surface area, the summation is over all $N_\mathrm{w}$ water molecules and $\delta(x)$ is Dirac's delta function.  As illustrated in Fig.~\ref{fig:system}c, $a_i^{(\mathrm{S})}$ and $a_i^{(\mathrm{I})}$ denote the distances of the oxygen atom of molecule $i$ from the substrate surface and instantaneous interface, respectively. 

A complementary measure of interfacial structure is the water density fluctuations  given by,  
\begin{equation}
\left \langle \left(\delta N^{(\alpha)}(z)\right)^2 \right \rangle = \left \langle \left (N^{(\alpha)}(z) - \langle N^{(\alpha)}(z) \rangle \right )^2 \right \rangle,
\end{equation}
where $N^{(\alpha)}(z)$ is the number of water molecules in spherical probe volume with radius $\sigma_\mathrm{p}=3\mathrm{\AA}$ and center a distance $z$ from the $\alpha$th interface ($\alpha$ being either S or I), i.e.,
\begin{equation}
N^{(\alpha)}(z) = \sum_{i=1}^{N_\mathrm{w}} \Theta \left( \sigma_\mathrm{p}-\sqrt{x_i^2 + y_i^2 + (a_i^{(\alpha)}-z)^2}  \right),
\end{equation}
where $\Theta(d)$ is the Heaviside function, either 1 if $d\ge0$ or 0 if $d < 0$, and $x_i$ and $y_i$ are the Cartesian coordinates of molecule $i$. 

In the next two subsections we analyze $\rho^{(\alpha)}(z)$ and $\left \langle (\delta N^{(\alpha)}(z))^2 \right\rangle$ for the full and intrinsic interfaces at model hydrophobic and hydrophilic substrates.

\section{Results and Discussion} 

\subsection{Hydrophobic Substrates}
Changing the parameter $\lambda$ changes the strength of substrate-water attractions and thus changes the hydrophobicity of the substrate. Figure~\ref{fig:system}(b) shows how a water-drop contact angle, $\theta$, reflects these changes.  According to our simulations, over the range of $\lambda$ considered, $\cos(\theta)$ is approximately a linear function of this attractive-interaction parameter. 

\begin{figure}
\includegraphics[width=3.37in]{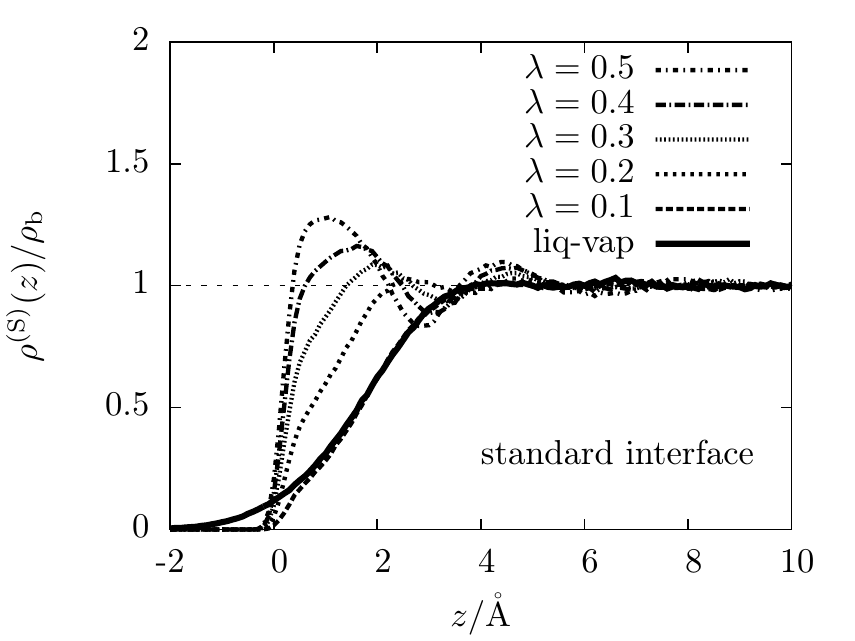}
\includegraphics[width=3.37in]{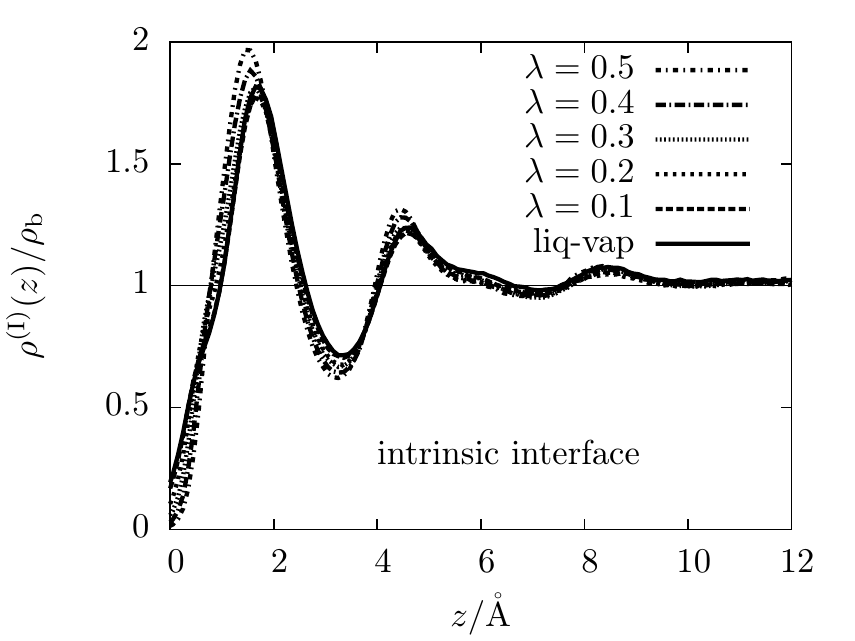}
\caption{The mean interfacial water density profile, $\rho^{(\alpha)}(z)$ is plotted for the standard interface ($\alpha=\mathrm{S}$) in the top panel and the intrinsic interface ($\alpha=\mathrm{I}$) in the bottom panel. Densities are normalized by the bulk liquid density $\rho_\mathrm{b}$.}
\label{fig:denprof}
\end{figure}

Figure~\ref{fig:denprof} shows the $\lambda$ dependnce of water density profile, $\rho^{(\alpha)}(z)$, computed for both the standard and the intrinsic interface.  For comparison, the average density profile of the free liquid-vapor interface is also shown.  The density profile for the standard interface, Fig.~\ref{fig:denprof}a, is notably sensitive to the relative hydrophobicity, exhibiting behavior ranging from a sigmoidal liquid-vapor-like profile at $\lambda=0.1$ to an oscillating profile indicative of molecular layering at $\lambda=0.5$. This sensitivity reflects the fact that the instantaneous interface is a soft collective variable. The fluctuations of the soft interface obscures the universal behavior of the intrinsic interface, whether in contact with a hydrophobic surface or a vapor phase.  Projecting out the spatial fluctuations of the soft liquid interface by focusing on the intrinsic interface, Fig.~\ref{fig:denprof}b shows a collapse of $\rho^{(\mathrm{I})}(z)$ onto a single curve. The intrinsic hydrophobic density profile exhibits pronounced molecular layering, and it is essentially indistinguishable from the density profile of the intrinsic liquid-vapor interface. 

\begin{figure}
\includegraphics[width=3.37in]{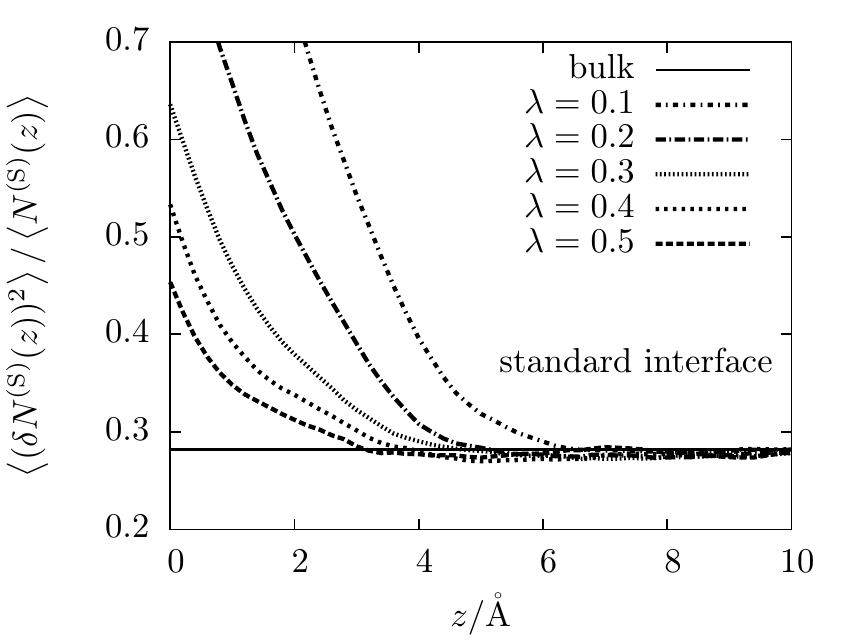}
\includegraphics[width=3.37in]{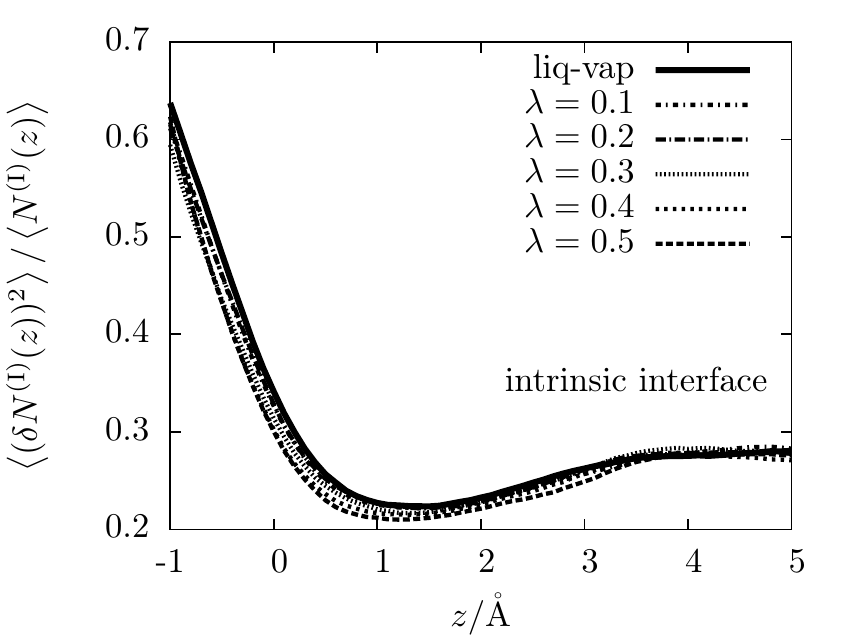}
\caption{The interfacial water density fluctuations in a $3\mathrm{\AA}$ spherical probe volume whose center is a distance $z$ from the position of the substrate surface, i.e. the standard interface (top), or the instantaneous liquid phase boundary, i.e. the intrinsic interface (bottom). The quantity in the denominator, $\left \langle N^{(\alpha)}(z) \right \rangle$ is the average number of water molecules in the same probe volume ($\alpha=\mathrm{S}$ or $\alpha=\mathrm{I}$ for the standard and intrinsic interface respectively). The horizontal line indicates the bulk value of the water density fluctuations.}
\label{fig:fluctprof}
\end{figure}

Fluctuations from the mean density profile illustrate the same point.  Specifically, Fig.~\ref{fig:fluctprof} shows our simulation results for $\left \langle (\delta N^{(\alpha)}(z))^2 \right \rangle$ for both $\alpha =$ S and $\alpha =$ I. For the standard substrate-water interface, density fluctuations are much larger near the substrate than in the bulk liquid. The magnitude and spatial variation of these fluctuations depend sensitively on $\lambda$.  In contrast, these fluctuations in reference to the instantaneous interface are at most weakly dependent on $\lambda$, and very much similar to those of the liquid-vapor interface.   

The $\lambda$ dependences of $\rho^{(\alpha)}(z)$ and $\left \langle (\delta N^{(\alpha)}(z))^2 \right \rangle$ indicate that the hydrophobic interface of water is indeed a liquid-vapor-like interface that is pinned to an attractive substrate. The attractions are weak in comparison to water-water hydrogen bonding, but strong enough to compete with the entropically driven capillary-wave motions of the phase boundary. The molecular structure of the standard interface therefore represents a convolution of the universal intrinsic molecular structure with a position distribution for the instantaneous liquid interface. The latter is substrate dependent and accounts for the observed sensitivity of interfacial structure on substrate hydrophobicity (such as that seen in Fig.~\ref{fig:denprof}a and Fig.~\ref{fig:fluctprof}a).

\begin{figure}
\includegraphics[width=3.37in]{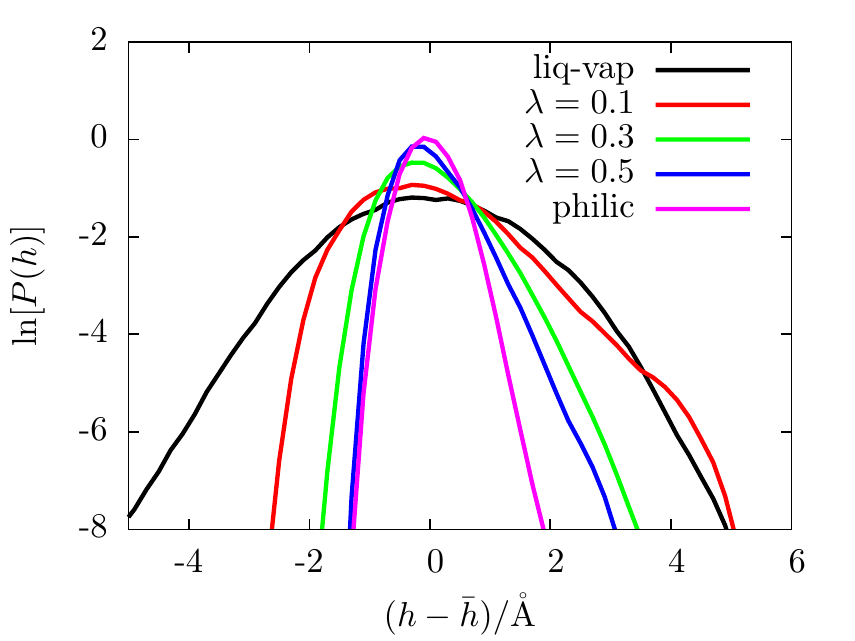}
\caption{The probability distribution governing height fluctuations of the instantaneous liquid interface.The distributions here are plotted relative to the average height of the interface $\bar{h}$.}
\label{fig:pofh}
\end{figure}

The distance between the instantaneous interface and the substrate -- the instantaneous interface height, $h$, defined in Fig.~\ref{fig:system}c) -- has a distribution of values.  This distribution, $P(h)$, provides another perspective on the story summarized in the previous paragraph.  Bear in mind that $P(h)$ is system-size dependent because of the relationship between capillary wave amplitude and wavelength~\cite{BW84,AJP11}. Nonetheless, for a series of identically sized systems, $P(h)$ provides qualitative insight into the statistics governing spatial fluctuations of water-substrate interfaces. Figure~\ref{fig:pofh} shows that $P(h)$ has large dependence on substrate identity. For a liquid-vapor interface, $P(h)$ is broad and roughly Gaussian, consistent with expectations from capillary wave theory. For the hydrophobic substrates $P(h)$ is narrower than the liquid-vapor case and asymmetric about the mean. The tails of $P(h)$ are truncated for fluctuations in the direction of the substrate ($h<\bar{h}$), which is a manifestation of substrate excluded volume.  In contrast, the tails are exaggerated for fluctuations of the interface into the bulk ($h>\bar{h}$). Those non-Gaussian fat tails are a signature of transient collective detachments of segments of the liquid interface from the weakly attractive substrate~\cite{DC00,APW09b,AJP10,AJP12} and hence are more pronounced for increasingly hydrophobic surfaces. Sensibly, therefore, the fat tails are most pronounced when $\lambda=0.1$ and become less so monotonically as $\lambda$ increases. 

\subsection{Hydrophilic Substrate}
\begin{figure}
\includegraphics[width=3.37in]{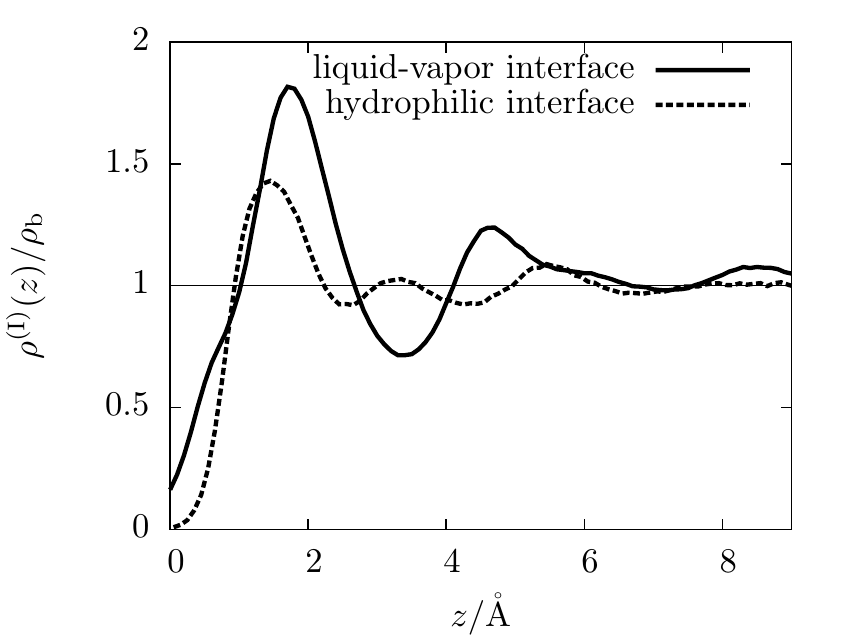}
\includegraphics[width=3.37in]{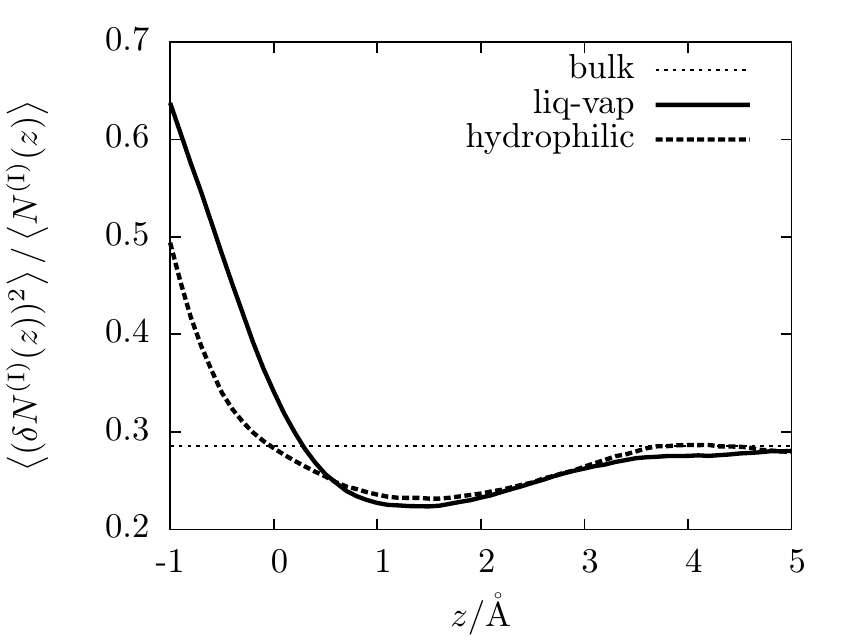}
\caption{(top) The mean interfacial density profile, $\rho^{(\mathrm{I})}(z)$, for the intrinsic hydrophilic and liquid-vapor interface are plotted with a solid and dashed line respectively. (bottom) Water density fluctuations, $\left \langle (\delta N^{(\mathrm{I})}(z))^2 \right \rangle$, for the intrinsic hydrophilic and liquid-vapor interface are plotted with a solid and dashed line respectively.}
\label{fig:denphilic}
\end{figure}

Unlike the picture we have drawn for hydrophobic surfaces, the behavior of water interfaces adjacent to hydrophilic surfaces are not liquid-vapor-like.  To see this fact, we follow the protocol in the previous subsection by computing $\rho^{(\mathrm{I})}(z)$ and $\left \langle (\delta N^{(\mathrm{I})}(z))^2 \right \rangle$ for the intrinsic interface between water and a model hydrophilic substrate (see Methods section for substrate details). The model hydrophilic substrate is locally polar and capable of forming favorable hydrogen bonds with water molecules in the liquid. As shown in Fig.~\ref{fig:denphilic}, unlike the hydrophobic case, at the intrinsic hydrophilic interface both $\rho^{(\mathrm{I})}(z)$ and $\left \langle (\delta N^{(\mathrm{I})}(z))^2 \right\rangle$ do not resemble their liquid-vapor counterparts. The solvent density $\rho^{(\mathrm{I})}(z)$ still exhibited molecular layering but with peak positions, and relative heights that are qualitatively different from that of a liquid-vapor interface. The $z$-dependence of $\left \langle (\delta N^{(\mathrm{I})}(z))^2 \right \rangle$ exhibits similar qualitative but different quantitative behavior than for that of a hydrophobic interface indicating that the solvation environment at a hydrophilic interface is fundamentally different than at a hydrophobic interface. In accordance with expectations that the liquid water interface interacts strongly with the hydrophilic substrate, the distribution of interface heights, $P(h)$, is both narrow and symmetric. 

\section{Acknowledgments}
Thanks to Shekhar Garde, David Limmer, Amish Patel, and Patrick Varilly for useful discussion. This research was enabled by the Helios Solar Energy Research Center and the CPIMS program, which are supported by the Director, Office of Science, Office of Basic Energy Sciences of the U.S. Department of Energy under Contract No. DE-AC02-05CH1123.

\nocite{*}
\bibliography{lambda}

\begin{thebibliography}{21}%
\makeatletter
\providecommand \@ifxundefined [1]{%
 \@ifx{#1\undefined}
}%
\providecommand \@ifnum [1]{%
 \ifnum #1\expandafter \@firstoftwo
 \else \expandafter \@secondoftwo
 \fi
}%
\providecommand \@ifx [1]{%
 \ifx #1\expandafter \@firstoftwo
 \else \expandafter \@secondoftwo
 \fi
}%
\providecommand \natexlab [1]{#1}%
\providecommand \enquote  [1]{``#1''}%
\providecommand \bibnamefont  [1]{#1}%
\providecommand \bibfnamefont [1]{#1}%
\providecommand \citenamefont [1]{#1}%
\providecommand \href@noop [0]{\@secondoftwo}%
\providecommand \href [0]{\begingroup \@sanitize@url \@href}%
\providecommand \@href[1]{\@@startlink{#1}\@@href}%
\providecommand \@@href[1]{\endgroup#1\@@endlink}%
\providecommand \@sanitize@url [0]{\catcode `\\12\catcode `\$12\catcode
  `\&12\catcode `\#12\catcode `\^12\catcode `\_12\catcode `\%12\relax}%
\providecommand \@@startlink[1]{}%
\providecommand \@@endlink[0]{}%
\providecommand \url  [0]{\begingroup\@sanitize@url \@url }%
\providecommand \@url [1]{\endgroup\@href {#1}{\urlprefix }}%
\providecommand \urlprefix  [0]{URL }%
\providecommand \Eprint [0]{\href }%
\providecommand \doibase [0]{http://dx.doi.org/}%
\providecommand \selectlanguage [0]{\@gobble}%
\providecommand \bibinfo  [0]{\@secondoftwo}%
\providecommand \bibfield  [0]{\@secondoftwo}%
\providecommand \translation [1]{[#1]}%
\providecommand \BibitemOpen [0]{}%
\providecommand \bibitemStop [0]{}%
\providecommand \bibitemNoStop [0]{.\EOS\space}%
\providecommand \EOS [0]{\spacefactor3000\relax}%
\providecommand \BibitemShut  [1]{\csname bibitem#1\endcsname}%
\let\auto@bib@innerbib\@empty
\bibitem [{\citenamefont {Stillinger}(1973)}]{FS73}%
  \BibitemOpen
  \bibfield  {author} {\bibinfo {author} {\bibfnamefont {F.~H.}\ \bibnamefont
  {Stillinger}},\ }\bibfield  {title} {\enquote {\bibinfo {title} {Structure in
  aqueous solutions of nonpolar solutes from the standpoint of scaled-particle
  theory},}\ }\href@noop {} {\bibfield  {journal} {\bibinfo  {journal} {J. Sol.
  Chem.}\ }\textbf {\bibinfo {volume} {2}},\ \bibinfo {pages} {141--158}
  (\bibinfo {year} {1973})}\BibitemShut {NoStop}%
\bibitem [{\citenamefont {Lum}, \citenamefont {Chandler},\ and\ \citenamefont
  {Weeks}(1999)}]{LCW}%
  \BibitemOpen
  \bibfield  {author} {\bibinfo {author} {\bibfnamefont {K.}~\bibnamefont
  {Lum}}, \bibinfo {author} {\bibfnamefont {D.}~\bibnamefont {Chandler}}, \
  and\ \bibinfo {author} {\bibfnamefont {J.~D.}\ \bibnamefont {Weeks}},\
  }\bibfield  {title} {\enquote {\bibinfo {title} {Hydrophobicity at small and
  large length scales},}\ }\href@noop {} {\bibfield  {journal} {\bibinfo
  {journal} {J. Phys. Chem. B}\ }\textbf {\bibinfo {volume} {103}},\ \bibinfo
  {pages} {4570--4577} (\bibinfo {year} {1999})}\BibitemShut {NoStop}%
\bibitem [{\citenamefont {Chandler}(2005)}]{DC05}%
  \BibitemOpen
  \bibfield  {author} {\bibinfo {author} {\bibfnamefont {D.}~\bibnamefont
  {Chandler}},\ }\bibfield  {title} {\enquote {\bibinfo {title} {Interfaces and
  the driving force of hydrophobic assembly},}\ }\href@noop {} {\bibfield
  {journal} {\bibinfo  {journal} {Nature}\ }\textbf {\bibinfo {volume} {437}},\
  \bibinfo {pages} {640--647} (\bibinfo {year} {2005})}\BibitemShut {NoStop}%
\bibitem [{\citenamefont {Bresme}, \citenamefont {Chac{\'o}n},\ and\
  \citenamefont {Tarazona}(2008)}]{PT08}%
  \BibitemOpen
  \bibfield  {author} {\bibinfo {author} {\bibfnamefont {F.}~\bibnamefont
  {Bresme}}, \bibinfo {author} {\bibfnamefont {E.}~\bibnamefont {Chac{\'o}n}},
  \ and\ \bibinfo {author} {\bibfnamefont {P.}~\bibnamefont {Tarazona}},\
  }\bibfield  {title} {\enquote {\bibinfo {title} {Molecular dynamics
  investigation of the intrinsic structure of water--fluid interfaces via the
  intrinsic sampling method},}\ }\href@noop {} {\bibfield  {journal} {\bibinfo
  {journal} {Phys. Chem. Chem. Phys.}\ }\textbf {\bibinfo {volume} {10}},\
  \bibinfo {pages} {4704--4715} (\bibinfo {year} {2008})}\BibitemShut {NoStop}%
\bibitem [{\citenamefont {Ashbaugh}\ and\ \citenamefont
  {Paulaitis}(2001)}]{MEP01}%
  \BibitemOpen
  \bibfield  {author} {\bibinfo {author} {\bibfnamefont {H.~S.}\ \bibnamefont
  {Ashbaugh}}\ and\ \bibinfo {author} {\bibfnamefont {M.~E.}\ \bibnamefont
  {Paulaitis}},\ }\bibfield  {title} {\enquote {\bibinfo {title} {Effect of
  solute size and solute-water attractive interactions on hydration water
  structure around hydrophobic solutes},}\ }\href@noop {} {\bibfield  {journal}
  {\bibinfo  {journal} {J. Am. Chem. Soc.}\ }\textbf {\bibinfo {volume}
  {123}},\ \bibinfo {pages} {10721--10728} (\bibinfo {year}
  {2001})}\BibitemShut {NoStop}%
\bibitem [{\citenamefont {Huang}\ and\ \citenamefont {Chandler}(2002)}]{DC02}%
  \BibitemOpen
  \bibfield  {author} {\bibinfo {author} {\bibfnamefont {D.~M.}\ \bibnamefont
  {Huang}}\ and\ \bibinfo {author} {\bibfnamefont {D.}~\bibnamefont
  {Chandler}},\ }\bibfield  {title} {\enquote {\bibinfo {title} {The
  hydrophobic effect and the influence of solute-solvent attractions},}\
  }\href@noop {} {\bibfield  {journal} {\bibinfo  {journal} {J. Phys. Chem. B}\
  }\textbf {\bibinfo {volume} {106}},\ \bibinfo {pages} {2047--2053} (\bibinfo
  {year} {2002})}\BibitemShut {NoStop}%
\bibitem [{\citenamefont {Zhou}\ \emph {et~al.}(2004)\citenamefont {Zhou},
  \citenamefont {Huang}, \citenamefont {Margulis},\ and\ \citenamefont
  {Berne}}]{BJB04}%
  \BibitemOpen
  \bibfield  {author} {\bibinfo {author} {\bibfnamefont {R.}~\bibnamefont
  {Zhou}}, \bibinfo {author} {\bibfnamefont {X.}~\bibnamefont {Huang}},
  \bibinfo {author} {\bibfnamefont {C.~J.}\ \bibnamefont {Margulis}}, \ and\
  \bibinfo {author} {\bibfnamefont {B.~J.}\ \bibnamefont {Berne}},\ }\bibfield
  {title} {\enquote {\bibinfo {title} {Hydrophobic collapse in multidomain
  protein folding},}\ }\href@noop {} {\bibfield  {journal} {\bibinfo  {journal}
  {Science}\ }\textbf {\bibinfo {volume} {305}},\ \bibinfo {pages} {1605--1609}
  (\bibinfo {year} {2004})}\BibitemShut {NoStop}%
\bibitem [{\citenamefont {Choudhury?}\ and\ \citenamefont
  {Montgomery~Pettitt}(2005)}]{MP05}%
  \BibitemOpen
  \bibfield  {author} {\bibinfo {author} {\bibfnamefont {N.}~\bibnamefont
  {Choudhury?}}\ and\ \bibinfo {author} {\bibfnamefont {B.}~\bibnamefont
  {Montgomery~Pettitt}},\ }\bibfield  {title} {\enquote {\bibinfo {title}
  {Local density profiles are coupled to solute size and attractive potential
  for nanoscopic hydrophobic solutes},}\ }\href@noop {} {\bibfield  {journal}
  {\bibinfo  {journal} {Mol. Sim.}\ }\textbf {\bibinfo {volume} {31}},\
  \bibinfo {pages} {457--463} (\bibinfo {year} {2005})}\BibitemShut {NoStop}%
\bibitem [{\citenamefont {Maibaum}\ and\ \citenamefont
  {Chandler}(2007)}]{LM07}%
  \BibitemOpen
  \bibfield  {author} {\bibinfo {author} {\bibfnamefont {L.}~\bibnamefont
  {Maibaum}}\ and\ \bibinfo {author} {\bibfnamefont {D.}~\bibnamefont
  {Chandler}},\ }\bibfield  {title} {\enquote {\bibinfo {title} {Segue between
  favorable and unfavorable solvation},}\ }\href@noop {} {\bibfield  {journal}
  {\bibinfo  {journal} {J. Phys. Chem. B}\ }\textbf {\bibinfo {volume} {111}},\
  \bibinfo {pages} {9025--9030} (\bibinfo {year} {2007})}\BibitemShut {NoStop}%
\bibitem [{\citenamefont {Godawat}, \citenamefont {Jamadagni},\ and\
  \citenamefont {Garde}(2009)}]{SG09}%
  \BibitemOpen
  \bibfield  {author} {\bibinfo {author} {\bibfnamefont {R.}~\bibnamefont
  {Godawat}}, \bibinfo {author} {\bibfnamefont {S.~N.}\ \bibnamefont
  {Jamadagni}}, \ and\ \bibinfo {author} {\bibfnamefont {S.}~\bibnamefont
  {Garde}},\ }\bibfield  {title} {\enquote {\bibinfo {title} {Characterizing
  hydrophobicity of interfaces by using cavity formation, solute binding, and
  water correlations},}\ }\href@noop {} {\bibfield  {journal} {\bibinfo
  {journal} {Proc. Nat. Acad. Sci.}\ }\textbf {\bibinfo {volume} {106}},\
  \bibinfo {pages} {15119--15124} (\bibinfo {year} {2009})}\BibitemShut
  {NoStop}%
\bibitem [{\citenamefont {Jamadagni}, \citenamefont {Godawat},\ and\
  \citenamefont {Garde}(2011)}]{SG11}%
  \BibitemOpen
  \bibfield  {author} {\bibinfo {author} {\bibfnamefont {S.~N.}\ \bibnamefont
  {Jamadagni}}, \bibinfo {author} {\bibfnamefont {R.}~\bibnamefont {Godawat}},
  \ and\ \bibinfo {author} {\bibfnamefont {S.}~\bibnamefont {Garde}},\
  }\bibfield  {title} {\enquote {\bibinfo {title} {Hydrophobicity of proteins
  and interfaces: Insights from density fluctuations},}\ }\href@noop {}
  {\bibfield  {journal} {\bibinfo  {journal} {Annu. Rev. Chem. Biomol. Eng.}\
  }\textbf {\bibinfo {volume} {2}},\ \bibinfo {pages} {147--171} (\bibinfo
  {year} {2011})}\BibitemShut {NoStop}%
\bibitem [{\citenamefont {Bandyopadhyay}\ and\ \citenamefont
  {Choudhury}(2012)}]{NC12}%
  \BibitemOpen
  \bibfield  {author} {\bibinfo {author} {\bibfnamefont {D.}~\bibnamefont
  {Bandyopadhyay}}\ and\ \bibinfo {author} {\bibfnamefont {N.}~\bibnamefont
  {Choudhury}},\ }\bibfield  {title} {\enquote {\bibinfo {title}
  {Characterizing hydrophobicity at the nanoscale: A molecular dynamics
  simulation study},}\ }\href@noop {} {\bibfield  {journal} {\bibinfo
  {journal} {J. Chem. Phys.}\ }\textbf {\bibinfo {volume} {136}},\ \bibinfo
  {pages} {224505} (\bibinfo {year} {2012})}\BibitemShut {NoStop}%
\bibitem [{\citenamefont {Willard}\ and\ \citenamefont
  {Chandler}(2010)}]{APW10}%
  \BibitemOpen
  \bibfield  {author} {\bibinfo {author} {\bibfnamefont {A.~P.}\ \bibnamefont
  {Willard}}\ and\ \bibinfo {author} {\bibfnamefont {D.}~\bibnamefont
  {Chandler}},\ }\bibfield  {title} {\enquote {\bibinfo {title} {Instantaneous
  liquid interfaces},}\ }\href@noop {} {\bibfield  {journal} {\bibinfo
  {journal} {J. Phys. Chem. B}\ }\textbf {\bibinfo {volume} {114}},\ \bibinfo
  {pages} {1954--1958} (\bibinfo {year} {2010})}\BibitemShut {NoStop}%
\bibitem [{\citenamefont {Berendsen}, \citenamefont {Grigera},\ and\
  \citenamefont {Straatsma}(1987)}]{SPCE}%
  \BibitemOpen
  \bibfield  {author} {\bibinfo {author} {\bibfnamefont {H.}~\bibnamefont
  {Berendsen}}, \bibinfo {author} {\bibfnamefont {J.}~\bibnamefont {Grigera}},
  \ and\ \bibinfo {author} {\bibfnamefont {T.}~\bibnamefont {Straatsma}},\
  }\bibfield  {title} {\enquote {\bibinfo {title} {The missing term in
  effective pair potentials},}\ }\href@noop {} {\bibfield  {journal} {\bibinfo
  {journal} {J. Phys. Chem.}\ }\textbf {\bibinfo {volume} {91}},\ \bibinfo
  {pages} {6269--6271} (\bibinfo {year} {1987})}\BibitemShut {NoStop}%
\bibitem [{\citenamefont {Weeks}, \citenamefont {Chandler},\ and\ \citenamefont
  {Andersen}(1971)}]{WCA}%
  \BibitemOpen
  \bibfield  {author} {\bibinfo {author} {\bibfnamefont {J.~D.}\ \bibnamefont
  {Weeks}}, \bibinfo {author} {\bibfnamefont {D.}~\bibnamefont {Chandler}}, \
  and\ \bibinfo {author} {\bibfnamefont {H.~C.}\ \bibnamefont {Andersen}},\
  }\bibfield  {title} {\enquote {\bibinfo {title} {Role of repulsive forces in
  determining the equilibrium structure of simple liquids},}\ }\href@noop {}
  {\bibfield  {journal} {\bibinfo  {journal} {J. Chem. Phys.}\ }\textbf
  {\bibinfo {volume} {54}},\ \bibinfo {pages} {5237--5247} (\bibinfo {year}
  {1971})}\BibitemShut {NoStop}%
\bibitem [{\citenamefont {Patel}, \citenamefont {Varilly},\ and\ \citenamefont
  {Chandler}(2010)}]{AJP10}%
  \BibitemOpen
  \bibfield  {author} {\bibinfo {author} {\bibfnamefont {A.~J.}\ \bibnamefont
  {Patel}}, \bibinfo {author} {\bibfnamefont {P.}~\bibnamefont {Varilly}}, \
  and\ \bibinfo {author} {\bibfnamefont {D.}~\bibnamefont {Chandler}},\
  }\bibfield  {title} {\enquote {\bibinfo {title} {Fluctuations of water near
  extended hydrophobic and hydrophilic surfaces},}\ }\href@noop {} {\bibfield
  {journal} {\bibinfo  {journal} {J. Phys. Chem. B}\ }\textbf {\bibinfo
  {volume} {114}},\ \bibinfo {pages} {1632--1637} (\bibinfo {year}
  {2010})}\BibitemShut {NoStop}%
\bibitem [{\citenamefont {Rowlinson}\ and\ \citenamefont {Widom}(2013)}]{BW84}%
  \BibitemOpen
  \bibfield  {author} {\bibinfo {author} {\bibfnamefont {J.~S.}\ \bibnamefont
  {Rowlinson}}\ and\ \bibinfo {author} {\bibfnamefont {B.}~\bibnamefont
  {Widom}},\ }\href@noop {} {\emph {\bibinfo {title} {Molecular theory of
  capillarity}}}\ (\bibinfo  {publisher} {Courier Dover Publications},\
  \bibinfo {year} {2013})\BibitemShut {NoStop}%
\bibitem [{\citenamefont {Patel}\ \emph {et~al.}(2011)\citenamefont {Patel},
  \citenamefont {Varilly}, \citenamefont {Jamadagni}, \citenamefont {Acharya},
  \citenamefont {Garde},\ and\ \citenamefont {Chandler}}]{AJP11}%
  \BibitemOpen
  \bibfield  {author} {\bibinfo {author} {\bibfnamefont {A.~J.}\ \bibnamefont
  {Patel}}, \bibinfo {author} {\bibfnamefont {P.}~\bibnamefont {Varilly}},
  \bibinfo {author} {\bibfnamefont {S.~N.}\ \bibnamefont {Jamadagni}}, \bibinfo
  {author} {\bibfnamefont {H.}~\bibnamefont {Acharya}}, \bibinfo {author}
  {\bibfnamefont {S.}~\bibnamefont {Garde}}, \ and\ \bibinfo {author}
  {\bibfnamefont {D.}~\bibnamefont {Chandler}},\ }\bibfield  {title} {\enquote
  {\bibinfo {title} {Extended surfaces modulate hydrophobic interactions of
  neighboring solutes},}\ }\href@noop {} {\bibfield  {journal} {\bibinfo
  {journal} {Proc. Nat. Acad. Sci.}\ }\textbf {\bibinfo {volume} {108}},\
  \bibinfo {pages} {17678--17683} (\bibinfo {year} {2011})}\BibitemShut
  {NoStop}%
\bibitem [{\citenamefont {Huang}\ and\ \citenamefont {Chandler}(2000)}]{DC00}%
  \BibitemOpen
  \bibfield  {author} {\bibinfo {author} {\bibfnamefont {D.~M.}\ \bibnamefont
  {Huang}}\ and\ \bibinfo {author} {\bibfnamefont {D.}~\bibnamefont
  {Chandler}},\ }\bibfield  {title} {\enquote {\bibinfo {title} {Cavity
  formation and the drying transition in the lennard-jones fluid},}\
  }\href@noop {} {\bibfield  {journal} {\bibinfo  {journal} {Phys. Rev. E}\
  }\textbf {\bibinfo {volume} {61}},\ \bibinfo {pages} {1501} (\bibinfo {year}
  {2000})}\BibitemShut {NoStop}%
\bibitem [{\citenamefont {Willard}\ and\ \citenamefont
  {Chandler}(2009)}]{APW09b}%
  \BibitemOpen
  \bibfield  {author} {\bibinfo {author} {\bibfnamefont {A.~P.}\ \bibnamefont
  {Willard}}\ and\ \bibinfo {author} {\bibfnamefont {D.}~\bibnamefont
  {Chandler}},\ }\bibfield  {title} {\enquote {\bibinfo {title} {Coarse-grained
  modeling of the interface between water and heterogeneous surfaces},}\
  }\href@noop {} {\bibfield  {journal} {\bibinfo  {journal} {Faraday
  Discussions}\ }\textbf {\bibinfo {volume} {141}},\ \bibinfo {pages}
  {209--220} (\bibinfo {year} {2009})}\BibitemShut {NoStop}%
\bibitem [{\citenamefont {Patel}\ \emph {et~al.}(2012)\citenamefont {Patel},
  \citenamefont {Varilly}, \citenamefont {Jamadagni}, \citenamefont {Hagan},
  \citenamefont {Chandler},\ and\ \citenamefont {Garde}}]{AJP12}%
  \BibitemOpen
  \bibfield  {author} {\bibinfo {author} {\bibfnamefont {A.~J.}\ \bibnamefont
  {Patel}}, \bibinfo {author} {\bibfnamefont {P.}~\bibnamefont {Varilly}},
  \bibinfo {author} {\bibfnamefont {S.~N.}\ \bibnamefont {Jamadagni}}, \bibinfo
  {author} {\bibfnamefont {M.~F.}\ \bibnamefont {Hagan}}, \bibinfo {author}
  {\bibfnamefont {D.}~\bibnamefont {Chandler}}, \ and\ \bibinfo {author}
  {\bibfnamefont {S.}~\bibnamefont {Garde}},\ }\bibfield  {title} {\enquote
  {\bibinfo {title} {Sitting at the edge: How biomolecules use hydrophobicity
  to tune their interactions and function},}\ }\href@noop {} {\bibfield
  {journal} {\bibinfo  {journal} {J. Phys. Chem. B}\ }\textbf {\bibinfo
  {volume} {116}},\ \bibinfo {pages} {2498--2503} (\bibinfo {year}
  {2012})}\BibitemShut {NoStop}%
\end{thebibliography}%

\end{document}